# Collective Variables Based on Multipole Expansion of Ewald Summation for Crystallization

*YaoKun Lei[1], MaoDong Li[1], and Yi Isaac Yang[1,*]*

[1] Institute of Systems and Physical Biology, Shenzhen Bay Laboratory, Shenzhen 518132, China

* To whom correspondence should be addressed: yangyi@szbl.ac.cn

**Abstract**
Crystallization, a fundamental phase transition process governing material formation in natural and industrial contexts, involves the spontaneous emergence of long-range structural order from disordered phases. This long-range periodicity involves spatial and molecular orientation order. Molecular dynamics (MD) simulations of crystallization require collective variables (CVs) that accurately distinguish this long-range periodicity. Existing CVs based on local descriptors (e.g., bond-orientational order) often lack transferability across crystal structures. To address this, we propose a unified CV framework derived from the multipole expansion of Ewald summation—a mathematical formalism bridging X-ray diffraction (XRD) principles and electrostatic energy computation in MD. By projecting atomic configurations onto a basis of spherical harmonics (complete for angular function representation), our CV achieves high-fidelity encoding of both translational and orientational order. Metadynamics simulations demonstrate that this CV drives efficient sampling of polymorphic pathways for known crystals and predicts stable phases even without crystal structures. This approach shows potential as a transferable platform for *ab initio* crystal structure prediction.

## 1. Introduction

Crystallization is a fundamental phase transition process that governs material formation across scientific, industrial, and natural contexts by establishing long-range periodic order in atomic/molecular spatial arrangements and orientations. This structural order underpins critical applications such as pharmaceutical purification—where it ensures the bioavailability of active ingredients (e.g., antibiotics, insulin) by controlling polymorphism and crystal morphology—and the synthesis of functional materials like semiconductors. Molecular dynamics (MD) simulations of crystallization require **collective variables (CVs)** capable of quantifying the emergence of translational symmetry and molecular orientational order to enable efficient sampling of nucleation pathways. While classical CVs like Steinhardt bond-order parameters[1, 2] or entropy-enthalpy[3, 4] metrics have been proposed, they face significant limitations:

1. **Orientational entropy descriptors** depend critically on user-defined angular cutoffs, necessitating prior knowledge of crystal symmetry and introducing subjective bias for unknown systems.
2. **X-ray diffraction (XRD)-inspired CVs** leverage peak intensities to reflect reciprocal-space periodicity but reduce molecules to point masses[5-8], discarding orientational information and requiring explicit reference structures for implementation.

To overcome these constraints, we propose a **transferable CV framework** grounded in the multipole expansion of Ewald summation—a mathematical formalism bridging electrostatic energy computation and scattering theory. This approach exploits three fundamental principles:

1. **Fourier duality**: Ewald summation and XRD both decompose periodicity via Fourier transforms, enabling reciprocal-space representation of lattice symmetry without predefined references.
2. **Multipole completeness**: Atomic configurations projected onto spherical harmonics (complete for angular functions) and Fourier basis (complete for periodic systems) encode *both* positional and orientational order.
3. **Local-global coupling**: Long-range crystallinity emerges from local interparticle correlations, making multipole moments transferable across crystal classes.

Our CV design offers distinct advantages:

- **Unified reciprocal-space representation**: Crystallinity metrics derive from summed reciprocal-space vectors, consolidating multiple symmetry indicators into a differentiable scalar function.
- **Orientation-aware multipoles**: Molecular orientation is preserved via multipole tensor, avoiding point-mass approximations.
- **Reference-free applicability**: Basis-set completeness ensures structural discrimination without prior crystal templates.

Metadynamics simulations demonstrate that this CV drives efficient sampling of polymorphic pathways for known crystals and predicts stable phases for systems lacking structural priors. By rigorously connecting electrostatic theory, scattering physics, and enhanced sampling, this framework shows potential as a transferable platform for *ab initio* crystal structure prediction.

## 2. Method

The XRD intensity in reciprocal space $\boldsymbol{Q}$ (reciprocal/scattering vector) derives from the structure factor $S(\boldsymbol{Q})$:

$$I(\boldsymbol{Q}) = \sum_i \sum_j f_i(\boldsymbol{Q}) f_j(\boldsymbol{Q}) \exp(-\boldsymbol{Q} \cdot \boldsymbol{r}_{ij}) = S(\boldsymbol{Q}) * S(-\boldsymbol{Q})$$

$$S(\boldsymbol{Q}) = \sum_i f_i(\boldsymbol{Q}) \exp(-\boldsymbol{Q} \cdot \boldsymbol{r}_i)$$

Here, $f_i(\boldsymbol{Q})$ is the scattering factor and $\boldsymbol{r}_i$ denotes position vector of i-th atom. In practical calculation, summations are restricted to molecular centers of mass, neglecting **molecular orientational order**.

Classical force field, compute long-range electrostatic energy in periodic conditions via Ewald summation:

$$U_{long} = \frac{2\pi}{V} \sum_{|\boldsymbol{Q}| \neq 0} \frac{1}{|\boldsymbol{Q}|^2} * \exp\left(-\frac{|\boldsymbol{Q}|^2}{4\alpha}\right) * S(\boldsymbol{Q}) * S(-\boldsymbol{Q})$$

$$S(\boldsymbol{Q}) = \sum_i q_i \exp(-\boldsymbol{Q} \cdot \boldsymbol{r}_i)$$

Here, V denotes the system volume, $q_i$ represents the partial atomic charge of the i-th atom in the classical force field, and $\alpha$ is the convergence parameter controlling the Gaussian width for reciprocal-space summation in Ewald methods. This formalism shares mathematical homology with XRD intensity calculations, revealing a unified Fourier-space description of periodicity.

To encode molecular orientation, each molecule j is represented by its multipole moments $\{\boldsymbol{M}_j^l\}$ (angular momentum 0≤l≤lmax) expanded about its center of mass $\boldsymbol{r}_j$:

$$U_{long} = \frac{2\pi}{V} \sum_{|\boldsymbol{Q}| \neq 0} \frac{1}{|\boldsymbol{Q}|^2} * \exp\left(-\frac{|\boldsymbol{Q}|^2}{4\alpha}\right) * S(\boldsymbol{Q}) * S(-\boldsymbol{Q})$$

The generalized structure factor becomes:

$$S(\boldsymbol{Q}) = \sum_j \tilde{L}_j(\boldsymbol{Q}) * \exp(-\boldsymbol{Q} \cdot \boldsymbol{r}_j)$$

where $\tilde{L}_j(\boldsymbol{Q})$ is the orientation-coupled kernel:

$$\tilde{L}_j(\boldsymbol{Q}) = \sum_{l=0}^{lmax} (2\pi i)^l * (\boldsymbol{M}_j^l \odot \boldsymbol{Q}^{\otimes l})$$

Here, $Q^{\otimes l} = \underbrace{Q \otimes Q \otimes \ldots \otimes Q}_{l\ times}$ denotes the l-fold tensor product, and $\odot$ the tensor contraction. Multipole moments $M_j^l$ preserve molecular orientation via spherical harmonics, overcoming the point-mass limitation of conventional XRD-inspired CVs. We derive CVs by omitting Ewald kernels ($\exp\left(-\frac{|Q|^2}{4\alpha}\right)$) and scaling factors, focusing on symmetry-sensitive terms:

$$CV^{(l_1, l_2)} = \sum_{Q \in \{Q\}} 0.5 * (S^{l_1}(Q) * S^{l_2}(-Q) + S^{l_2}(Q) * S^{l_1}(-Q))$$

with

$$S^{l_1}(Q) = \sum_j \left(M_j^{l_1} \odot Q^{\otimes l_1}\right) * \exp(-Q \cdot r_j)$$

The summation operates over a user-defined subset $\{Q\}$ of reciprocal-space vectors. $CV^{(l_1, l_2)}$ corresponds to the interaction between $l_1$ and $l_2$ order multipole and characterize distinct type of molecular order. For the time being, we implement the following three terms:

$CV^{(0,0)}$: encodes center-of-mass spatial periodicity.
$CV^{(1,1)}$: quantifies the **alignment** of molecular dipoles.
$CV^{(2,2)}$: describes the **nematic order** associated with molecular quadrupoles.

For neutral molecules, the monopole moment $M_j^0$ vanishes. To preserve mass-center periodicity information in the $CV^{(0,0)}$ calculation, we assign $M_j^0 = 1$ universally, reducing this term to a **density correlation function**. While arbitrary geometric tensors (e.g., vector products) could theoretically replace multipoles in this framework, multipole moments are physically preferred for two reasons:

1. **Local-global correspondence**: Long-range crystallinity emerges from localized electrostatic interactions, which multipoles inherently encode via their origin at molecular mass centers.
2. **Minimal representation**: Multipoles provide a non-redundant basis for anisotropic interactions, as their symmetric trace-free property eliminates spurious degrees of freedom.

The collective variables leverage dual complete basis sets:
- **Spherical harmonics**: Complete for angular functions on $\mathbb{S}^2$, enabling exact representation of molecular orientation.
- **Fourier basis**: Complete for periodic functions in $\mathbb{R}^3$, encoding long-range translational symmetry.

This dual completeness guarantees **reference-free structural discrimination**, as the basis sets span all possible symmetry representations without requiring prior crystal templates.

## 3. Results

The proposed CV is combined with metadynamics to simulate crystallization process. With known crystal structures, the contribution from each reciprocal vector is calculated. Only terms with distinguished intensity compared with liquid state is involved in CV calculation.

Benzene:
Biased force along $CV^{(2,2)}$ efficiently drive the benzene's backbone align to specific direction.

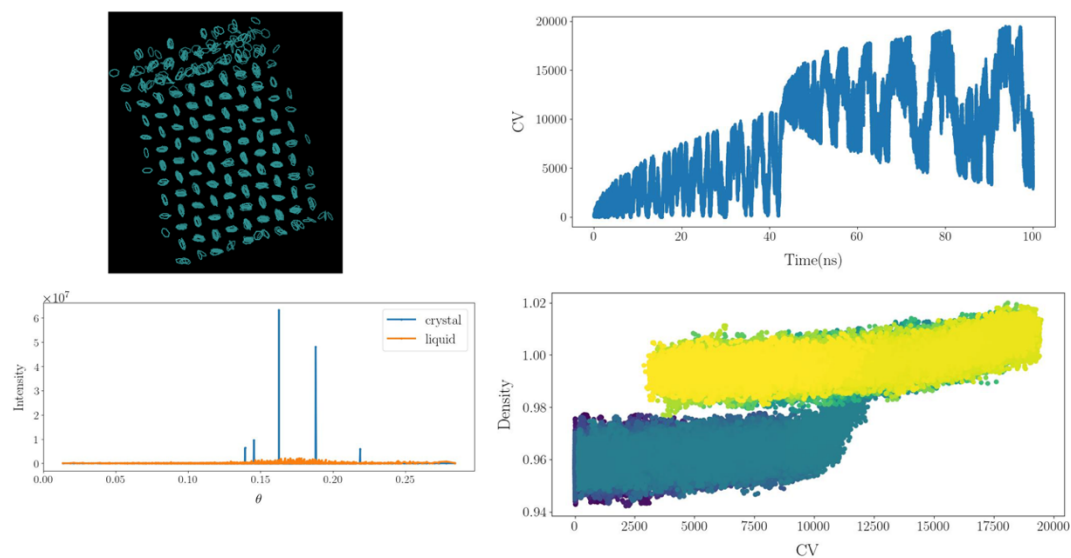

Figure 1 Upper left panel: simulated crystal structure of benzene. Upper right panel: Time evolution of $CV^{(2,2)}$. Lower left panel: Reciprocal-space contributions to $CV^{(2,2)}$ mapped to Bragg angles 2θ. Lower right panel: The density changes along $CV^{(2,2)}$.

Water:

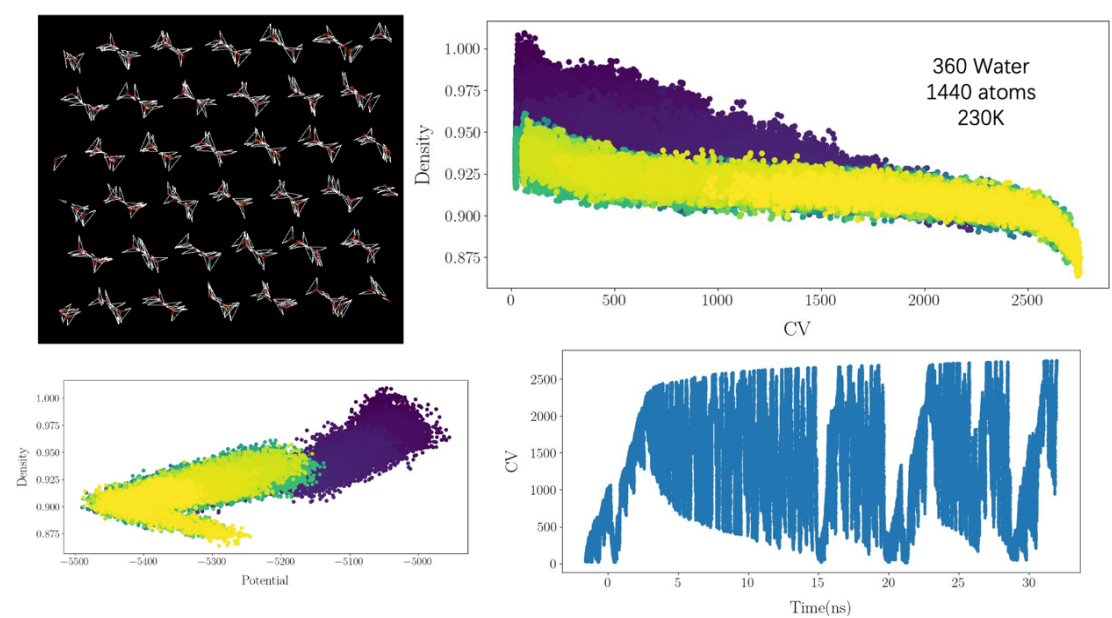

Figure 2 Upper left panel: simulated crystal structure of water. Upper right panel: The density changes along $CV^{(0,0)}$. Lower left panel: Sampled data plotted in density-potential space.

Urea:

There are two types of crystal structures for urea. Even though $CV^{(1,1)}$ can accelerate the alignment of dipole of urea, it can not differentiate two crystal structures with different orientations of plane containing carbonyl.

Urea exhibits multiple crystalline polymorphs under varying thermodynamic conditions. While the collective variable $CV^{(1,1)}$ accelerates dipole alignment during urea crystallization, it fails to distinguish polymorphs differing in **carbonyl-plane orientation** due to two key limitations:
  1. **Rotational invariance**: $CV^{(1,1)}$ quantifies net dipole alignment but lacks sensitivity to the spatial distribution of carbonyl groups relative to crystal axes.
  2. **Symmetry insensitivity**: Crystal 1 and Crystal 2 exhibit identical dipole magnitudes but distinct hydrogen-bonding networks that reorient carbonyl planes—information not captured by dipole-based CVs .

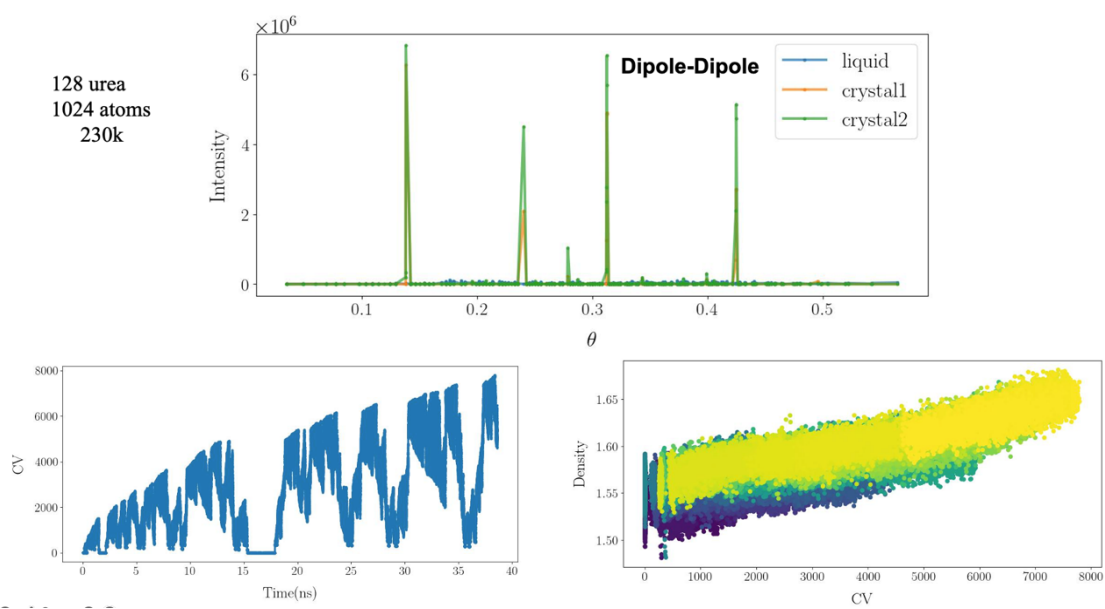

Figure 3 Top panel: Reciprocal-space contributions to $CV^{(1,1)}$ mapped to Bragg angles 2θ. Lower left panel: time evolution of $CV^{(1,1)}$. The density changes along $CV^{(1,1)}$.

$CV^{(2,2)}$ distinguishes between crystal structures exhibiting **different orientations of molecular planes containing carbonyl groups**.

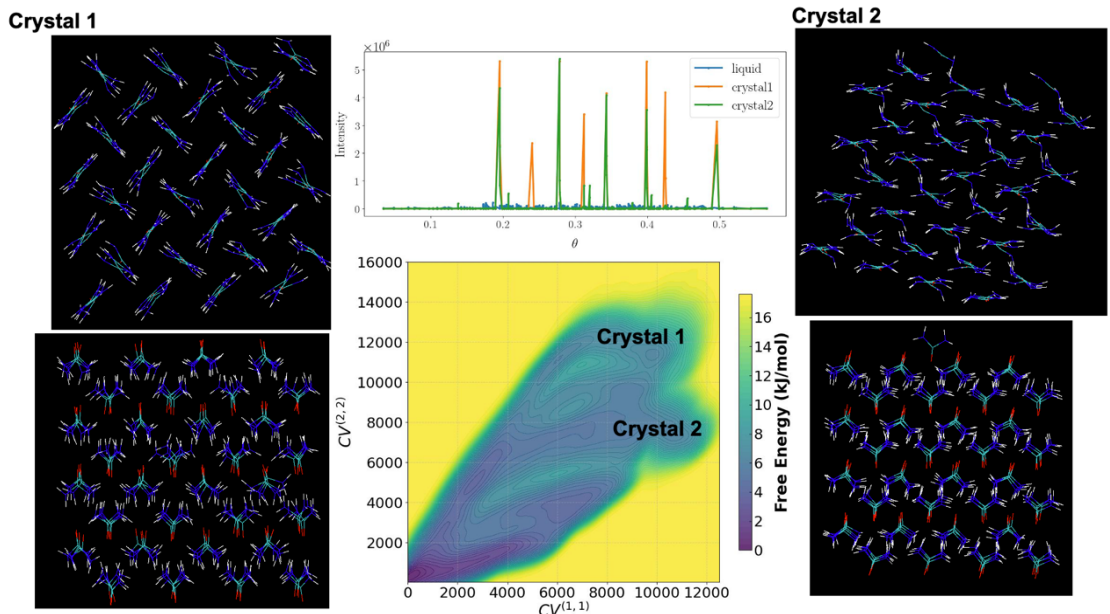

Figure 4 Left panel: Simulated crystal structures of Crystal 1 of urea. Right panel: Simulated crystal structures of Crystal 2. Upper middle panel: Reciprocal-space contributions to $CV^{(2,2)}$ mapped to Bragg angles 2θ. Lower middle panel: free energy plot in $CV^{(1,1)}$- $CV^{(2,2)}$ space (not reweighted).

Naphthalene:
When reciprocal-space peaks are predefined, $CV^{(2,2)}$ drives efficient structural sampling of crystal structures.

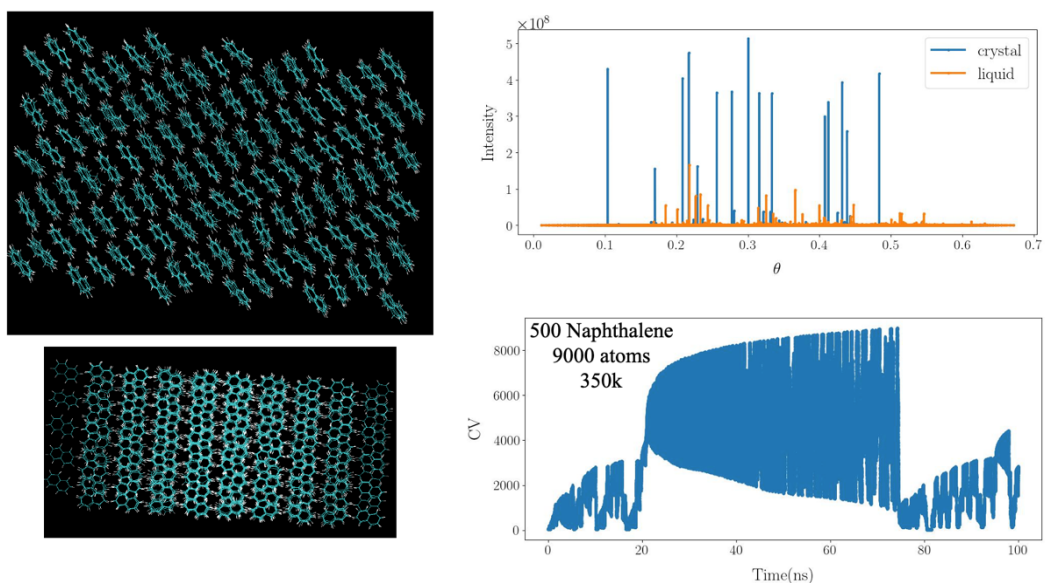

Figure 5 Left panel: Simulated crystal structures of naphthalene. Upper right panel: Reciprocal-space contributions to $CV^{(2,2)}$ mapped to Bragg angles 2θ. Lower right panel: time evolution of $CV^{(2,2)}$.

$CV^{(2,2)}$ with uniform distribution of reciprocal vectors can still sample crystal structures, demonstrating its potential for predicting naphthalene crystal.

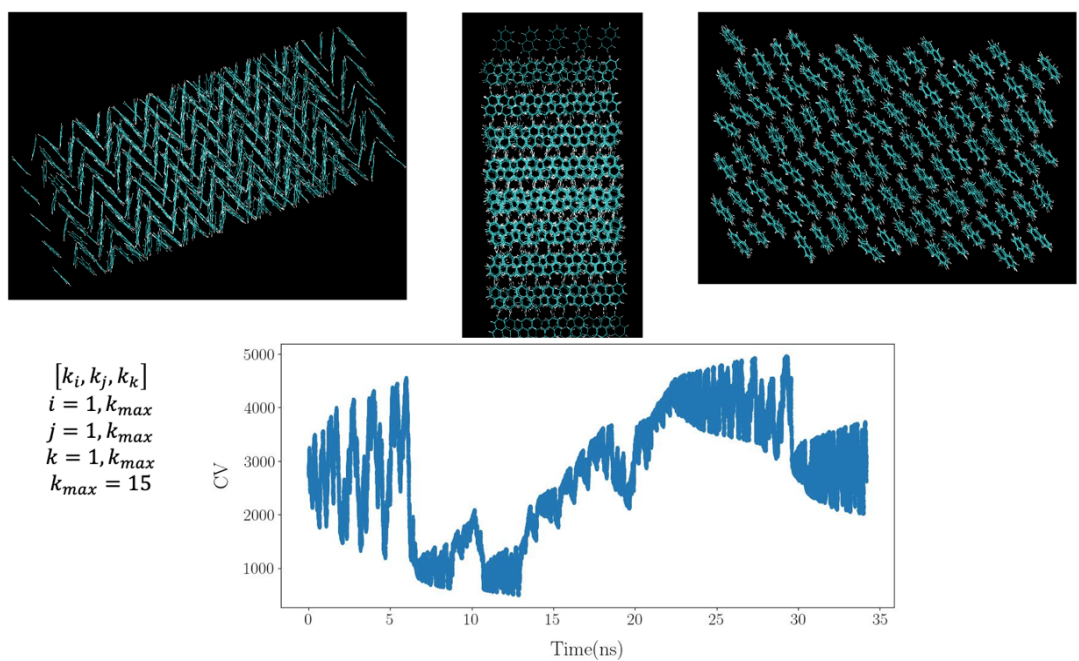

Figure 6 Top panel: Simulated crystal structures of naphthene. Lower left panel: reciprocal vectors used. Lower right panel: time evolution of $CV^{(2,2)}$.

## 4. Summary

We propose a **transferable CV framework** derived from the multipole expansion of Ewald summation, integrating spherical harmonics (for angular symmetry) with the Fourier basis (for translational periodicity). This CV drives efficient sampling of crystallization pathways **with expertise-free selection of crystallization descriptors** and **enables *ab initio* prediction of crystal phases in validated systems**.